\documentclass[12pt,a4paper]{article}
\usepackage[english]{babel}
\usepackage{amsmath}
\usepackage{mathrsfs}
\usepackage{graphicx}
\usepackage{amsfonts}
\usepackage{authblk}

\newcommand{\figref}[1]{Fig.~\ref{fig:#1}}

\renewcommand{\eqref}[1]{(\ref{eq:#1})}

\begin{document}
\title{Numerical evidence of Sinai diffusion of random-mass Dirac particles}

\author[1]{Silvia Palpacelli\thanks{silvia.palpacelli@hyperlean.eu}}
\author[2]{Sauro Succi\thanks{succi@iac.cnr.it}}
\affil[1]{Hyperlean S.r.l, Via Giuseppe Verdi 4, 60122, Ancona, Italy}
\affil[2]{Istituto Applicazioni Calcolo, CNR, via dei Taurini 19, 00185, Roma, Italy}


\maketitle

\begin{abstract}
We present quantum Lattice Boltzmann simulations of the Dirac equation
for quantum-relativistic particles with random mass.
By choosing zero-average random mass fluctuation, the simulations show
evidence of localization and ultra-slow Sinai diffusion, due to the
interference of oppositely propagating branches of the quantum wavefunction
which result from random sign changes of the mass around a zero-mean.
The present results indicate that the quantum lattice Boltzmann scheme may offer a viable
tool for the numerical simulation of quantum-relativistic transport phenomena in topological materials.
\end{abstract}

\section{Introduction}
Topological materials, such as insulators and superconductors, have attracted
enormous interest in the recent years, as a novel form of quantum matter
exhibiting a whole array of fascinating properties \cite{BEEN,MORI}.
In particular, {\it disordered} low-dimensional quantum materials
present an intriguing interplay between topology and localization, eventually leading to
the suppression of Anderson localization in the vicinity of a phase transition
between two distinct topological phases \cite{PWA,SACH}.
This motivates an intense study of the basic mechanisms which govern the
transport properties of low-dimensional quantum materials in the vicinity
of a topological phase transitions.
Recent work in this direction has highlighted an intriguing tendency to
ultra-retarded transport as described by  classical {\it Sinai diffusion}
in a random {\it force} field.
Such tendency is, at least at a first glance, quite surprising, for there is no reason a-priori
why transport phenomena in low-dimensional quantum disordered materials
should exhibit any analogy with a classical Langevin particle in a random force field.
Such an analogy has been analysed in a recent paper, in the context of
quantum transport in Majorana quantum wires \cite{KAMEN}, based on analytical
inspection of the one-dimensional Dirac equation for a free quantum-relativistic
particle with a zero-average random mass.
In this paper, we analyse a similar problem by means of numerical simulations
of the random-mass Dirac equation based on the quantum lattice Boltzmann method
\cite{QLB1}.
Our numerical results confirm strong localization at increasing noise strength,
and provide indications of ultra-slow Sinai diffusion as well.

\section{Sinai diffusion: from classical to quantum regimes}

In 1982 Sinai analyzed the classical motion of a stochastic particle moving
in a random {\it force} field \cite{SINAI},
\begin{eqnarray}
\label{LANGE}
\frac{dz}{dt} = p/m\\
\frac{dp}{dt} = -\gamma p + \tilde F
\end{eqnarray}
where $p=mv$ is the particle momentum, $\gamma$ the friction frequency and
$\tilde F$ a random force with zero mean and variance $D$.
Analytical treatment of the above Langevin equation, shows that diffusion proceeds
at ultra-slow rates, with a variance scaling as:
\begin{eqnarray}
\label{VAR}
<z^2> = \lambda^2 log^4(t)
\end{eqnarray}
$\lambda \propto 1/D$ being the localization length of the corresponding
probability distribution $p(z)$.
Such ultra-slow diffusion is basically due to the fact that a random force
translates into a relatively regular potential $V(z)$, whose barriers
are more effective than random ones in trapping the particle in
the corresponding local minima.
The first evidences that Sinai diffusion might be relevant to
relativistic quantum transport was provided by pioneering work
on one-dimensional random mass Dirac fermions and equivalent
random Ising chains \cite{BAL,SHANKAR}.

Indeed, the analysis of the Dirac equation with a random mass, such that
\begin{eqnarray}
<m(z)>=0\\
<m(z)m(z')> = 2 \lambda^{-1} \delta (z-z')
\end{eqnarray}
shows that the spectrum becomes not only gapless but also quantum critical, i.e.
it supports order-disorder transitions driven by purely
quantum fluctuations, with no contributions from thermal ones.
This can be heuristically understood by nothing that
the zero-energy (steady-state) Dirac equation is formally solved
by $\psi_{\pm}(z) \propto e^{\mp \int_{-\infty}^z m(z')dz'}$, so that, away
from null points, where $m(z^*)=0$, the critical solutions
decay like $e^{-\sqrt{|z-z^*|/\lambda}}$.
For small positive-energy solutions $E>0$, the behaviour remains similar,
and it can be shown that states separated in space by an overlap $d$
repel each other with an energy
$E = \hbar \omega \sim e^{-\sqrt{d/\lambda}}$, leading
to correlations between spatial and time scales of the order of
$d \sim \lambda ln^2(1/\omega)$.
By letting $t=1/\omega$, this gives precisely Sinai's diffusion.

\section{Numerical set-up}

Numerical solution of random-mass Dirac equations have been performed before \cite{YOSPRA}.
However, no special emphasis on connections between localization and
Sinai's diffusion have been provided.
In the sequel, we focus precisely on this aspect, by using the quantum lattice Boltzmann (QLB) method, to
be shortly described in the next section.

\subsection{The QLB model}

Let us recall the main ideas behind the quantum lattice Boltzmann scheme in one dimension.
Consider the Dirac equation in one dimension. Using the Majorana representation \cite{LL} and
projecting upon chiral eigenstates, the Dirac equation reads:
\begin{equation} \label{eq:dirac}
\begin{split}
\partial_t u + c \partial_z u & = \omega_c d,\\
\partial_t d - c \partial_z d & = - \omega_c u,
\end{split}
\end{equation}
where $u$ and $d$ are complex wave functions composing a two-component spinor $\psi = (u, d)^T$,
and $\omega_c = mc^2/\hbar$ is the Compton frequency.\\ As observed in Ref. \cite{QLB1}, Eq. \eqref{dirac}
is a discrete Boltzmann equation for a couple of complex wave functions $u$
and $d$. In particular, the propagation step consists of streaming $u$ and $d$ along the $z$-axis with
opposite speeds $\pm c$, while the collision step is performed according to the scattering
matrix defined by the right hand side of Eq. \eqref{dirac}.\\

The QLB scheme is obtained by integrating Eq. \eqref{dirac} along the characteristics of $u$ and
$d$ respectively and approximating the right hand side integral by using the trapezoidal rule.
Assuming $\Delta z = c \Delta t$, the following scheme is
obtained
\begin{equation} \label{eq:eq_QLB}
\begin{split}
\hat{u} - u &= \frac{\widetilde{m}}{2} (d+ \hat{d})\\
\hat{d} - d &= - \frac{\widetilde{m}}{2} (u+ \hat{u}),
\end{split}
\end{equation}
where $\hat{u}=u(z+\Delta z,t+\Delta t)$, $\hat{d}=d(z-\Delta z,t+\Delta t)$, $u=u(z,t)$, $d=d(z,t)$
and $\widetilde{m} = \omega_c \Delta t$ is the dimensionless Compton frequency.

The linear system of
Eq. \eqref{eq_QLB} is algebraically solved for $\hat{u}$ and $\hat{d}$ and yields the explicit scheme:
\begin{equation} \label{eq:real-QLB}
\begin{split}
\hat{u} &= a u + b d,\\
\hat{d} &= a d - b u,
\end{split}
\end{equation}
where
\begin{equation*}
a = (1-\Omega/4)/(1 + \Omega/4), \quad b = \widetilde{m}/(1 + \Omega/4),
\end{equation*}
with $\Omega = \widetilde{m}^2$.
Note that, since $|a|^2 + |b|^2 = 1$, the collision matrix is unitary, thus the method is unconditionally
stable and norm-preserving.\\
In natural units ($c=1$, $\hbar=1$), $\omega_c=m$ and $\widetilde{m} = m \Delta t$.

\section{Numerical Results}

The ability of the QLB method to solve the Dirac equation for particle with constant mass  has been
discussed before in the literature \cite{DELLAR1,DELLAR2}.
In this section we present numerical results of the propagation of a free Dirac particle with
random mass. To the best of our knowledge, this is the first time QLB is applied to such a problem.

\subsection{Simulations of a free Dirac particle with random mass}

\begin{figure}[htbp]
\centering
(a)\\
\includegraphics[scale=0.45]{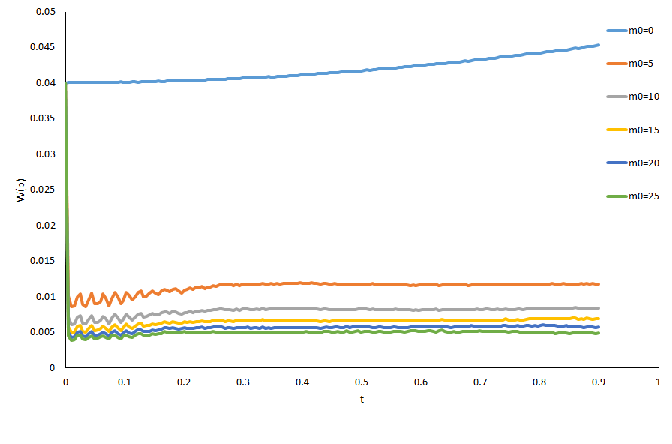}\\
(b)\\
\includegraphics[scale=0.45]{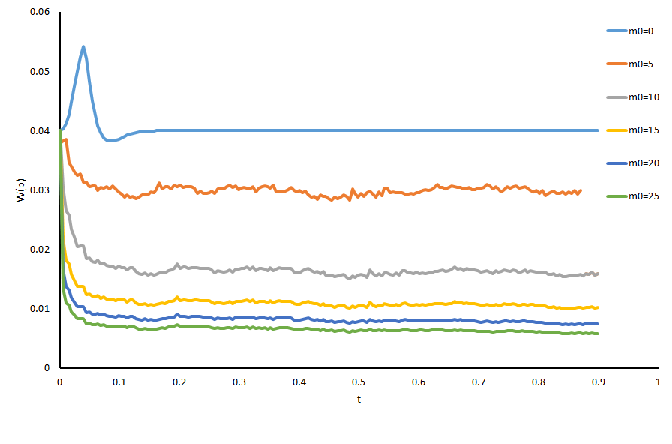}
\caption{$W[p(z)]$ as a function of time for $\overline{m}=500$ (a) and $\overline{m}=0$ (b) and for the following values of $m_0=0$, $5$, $10$, $15$, $20$ and $25$. In (b), for $m_0=0$, $W[p(z)]$ is constant apart from the initial peak which corresponds to the separation of the wavepacket "branches".
By increasing the noise amplitude $m_0$ leads to increasing localization.}
\label{fig:mass0_W}
\end{figure}

\begin{figure}[htbp]
\centering
(a)\\
\includegraphics[scale=0.45]{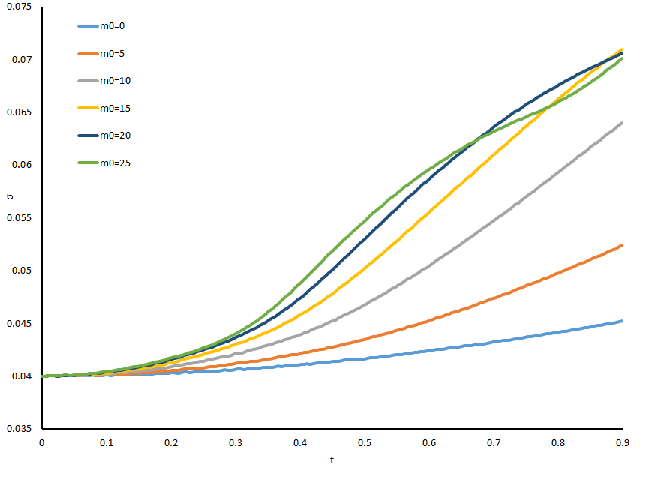}\\
(b)\\
\includegraphics[scale=0.45]{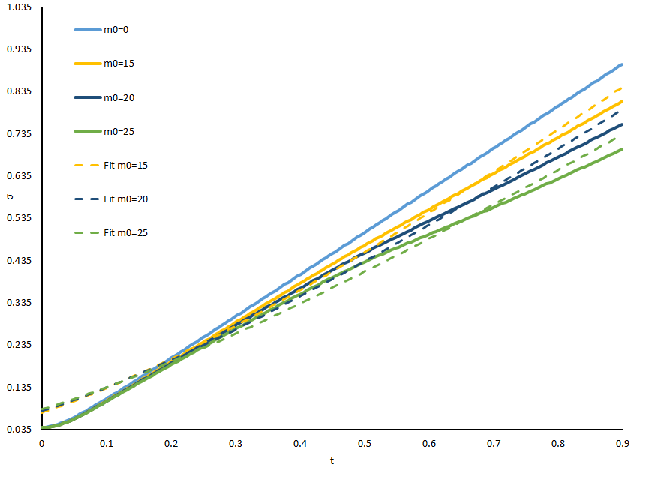}
\caption{Wave packet dispersion as a function of time computed as in Eq. \eqref{sigma} for $\overline{m}=500$ (a) and $\overline{m}=0$ (b) and for the following values of $m_0=0$, $5$, $10$, $15$, $20$ and $25$, in (b) curves for $m_0=5$ and $m_0=10$ are not shown in order to improve visual readability.
In (b), the dashed lines represent the power-logarithmic fit $f(t)=C ln^2(t+D)$ where $C$ and $D$ are varied in order to obtain the best fit for each value of $m_0$.
Although not perfect, the fit provides an indication of ultra-slow Sinai diffusion for $\overline{m}=0$.
The case $\overline{m}=500$, on the other hand provides no evidence of Sinai diffusion.
}
\label{fig:mass0_D}
\end{figure}

\begin{figure}[htbp]
\centering
(a) \hspace{5cm} (b) \\
\includegraphics[scale=0.35]{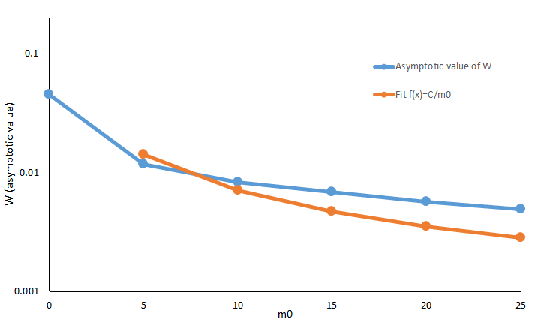}
\includegraphics[scale=0.35]{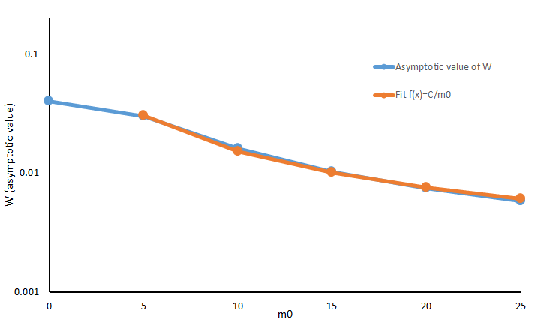}
\caption{Asymptotic value $W[p(z)]$ as a function of $m_0$, a fit with the function $f(x)=C/m_0$ is also shown with $C=0.0703$ and $C=0.151$ for $\overline{m}=500$ (a) and $\overline{m}=0$ (b).
The figure shows that for $\overline{m}=0$ the localization length scales inversely with the noise amplitude, following a sort of uncertainty relation $W m_0 \sim const$.}
\label{fig:Wasymptotic}
\end{figure}

In this section, the time-evolution of a Gaussian wave packet is computed by using a random mass
in order to inspect the effect of the random component on the wave packet localization.\\ In particular, the following initial condition is used
\begin{equation}
\label{eq:IC_Randmass}
  \psi(z,0)\equiv \begin{pmatrix} u \\ d \end{pmatrix} =N \exp(-z^2/4 \sigma^2 ) \begin{pmatrix} 1\\ -i \end{pmatrix}.
\end{equation}
Following Ref. \cite{YOSPRA}, we define a localization functional in order to "measure" the localization of a waveform in one spatial dimension
\begin{equation}
L[p(z)]:=\int_{-\infty}^{\infty} \lvert \sqrt{p(z)} \frac{d^2}{dz^2} \sqrt{p(z)}  \rvert dz
\end{equation}
where $p(z)=|\psi(z)|^2$.
An approximated localization width for $p(z)$ is defined by
\begin{equation}
\label{eq:W}
W[p(z)]:=\sqrt{C/L[p(z)]}
\end{equation}
with $C \sim 0.3789041452$ \cite{YOSPRA}.\\
We set $\sigma = 0.04$ and a Gaussian distributed random mass with average $<m> \equiv \overline{m}=500$ and $\overline{m}=0$ (in order to observe Sinai diffusion) and standard deviation $m_0=0$, $5$, $10$, $15$, $20$ and $25$, the spatial domain $[-1,1]$ is discretized with a lattice of size $N=2048$.
For each value of $m_0$, ensemble average over $100$ realizations has been performed.\\
In figures 1-5, the strong effect due to the random mass is shown, for each value of $m_0$ and
$\overline{m}=500$ (a) and $\overline{m}=0$ (b), respectively.
In particular, in \figref{mass0_W}, the localization function \eqref{W} is reported
for each value of $m_0$ showing an increasing localization at increasing values of $m_0$.\\
\figref{mass0_D} shows the wave packet dispersion as a function of time. The evolution of the wave packet dispersion is computed as follows
\begin{equation}
\label{eq:sigma}
\sigma(t) = \left( \int_{-\infty}^{+\infty} (z-<z>)^2 |\psi(z,t)|^2 dz \right)^{1/2},
\end{equation}
where
\begin{equation}
<z>=\int_{-\infty}^{+\infty} z |\psi(z,t)|^2 dz, \quad |\psi(z,t)|^2 = |u(z,t)|^2 + |d(z,t)|^2.
\end{equation}
In \figref{mass0_D} (b) for each curve a fit with
the function $f(t) = C \ln^2(t+D)$, is also reported, from which we can observe
that for $\overline{m}=0$, the wave packet dispersion follows a $ln^2(t)$ law, as expected
for the Sinai diffusion.\\
Moreover, in \figref{Wasymptotic}, the asymptotic value of $W[p(z)]$ is plotted as a function
of $m_0$ and compared with a fit against the function $f(m_0)=C/m_0$.
The best fit is obtained for $C=0.0703$ and $C=0.151$ for $\overline{m}=500$ (a) and $\overline{m}=0$ (b), respectively.

Finally, in \figref{mass0_psi} the evolution in time of $|\psi(z,t)|^2$ is shown for a
single realisation of the random mass, for $\overline{m}=500$ (a) and $\overline{m}=0$ (b), respectively.\\

From this figure, we observe that for the massive case $\overline{m}=500$, the up and downward moving
wave packets remain tightly coupled throughout, due to the strong inter-mixing due to the non-zero mass.
Thus, the two spinor components behave much like a single (non-relativistic) wavepacket, which gets increasingly
corrugated due to the random fluctuations. Note that since none of such fluctuations comes any close to the mean
value $\overline{m}=500$, the corrugations are comparatively weak and do not alter
the qualitative space-time pattern of the wavepacket.

The case with $\overline{m}=0$ shows a more intriguing story.
In the absence of noise, the up and down components of the spinor propagate
freely and independently along the $\pm z$ axis, respectively.
As noise is switched on around a zero-mass average, negative values of the mass eventually
occur, which spawn side-branches propagating in the opposite direction, i.e.
a negative-mass fluctuation of the up-mover along $+z$ spawns a down-mover side-branch, and viceversa.
Note indeed that the mass-inversion symmetry $ m \leftrightarrow -m$, as combined with the
exchange of the up-down spinors, $u \leftrightarrow d$, leaves the Dirac equation invariant.
The side-branches, in turn, spawn secondary side branches in a sort of recursive
cascade which gives rise to a "network" of correlations.
Such correlations progressively fill the ground between the two original
up and down movers, resulting in a highly corrugated and "networked" wave function.
Remarkably, the dispersion of such "networked" wavefunction obeys Sinai's diffusion scaling.

A word of caution is in order: the logarithmic fits of the wave packet dispersion are found
to be fairly sensitive to the numerical values of the fitting parameters $C$ and $D$.
Consequently, a statistically accurate assessment of the actual exponent of Sinai
ultra-low diffusion must await for further systematic studies.
Work along these lines is currently underway.
\begin{figure}[htbp]
\centering
$m_0=0$ \hspace{5cm} $m_0=5$\\ \vspace{0.15cm}
\includegraphics[scale=0.22]{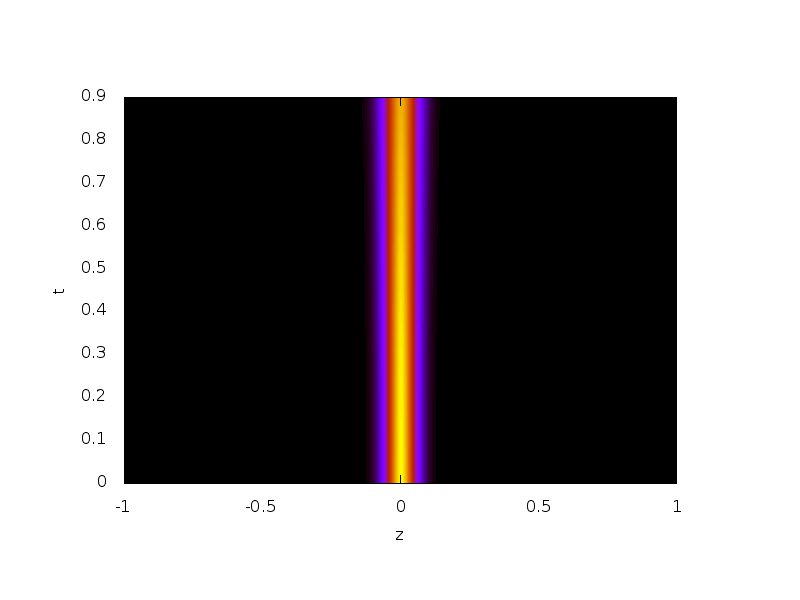}
\includegraphics[scale=0.22]{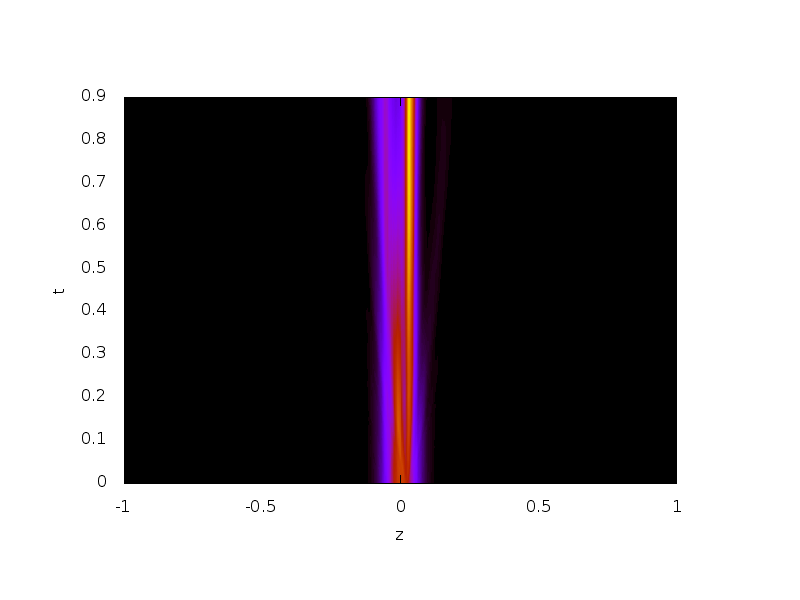}\\
$m_0=10$ \hspace{5cm} $m_0=15$\\ \vspace{0.15cm}
\includegraphics[scale=0.22]{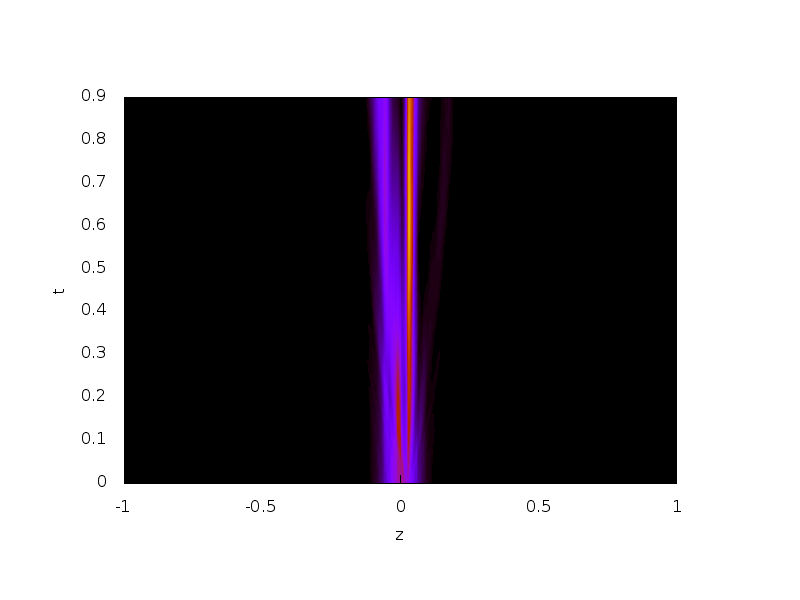}
\includegraphics[scale=0.22]{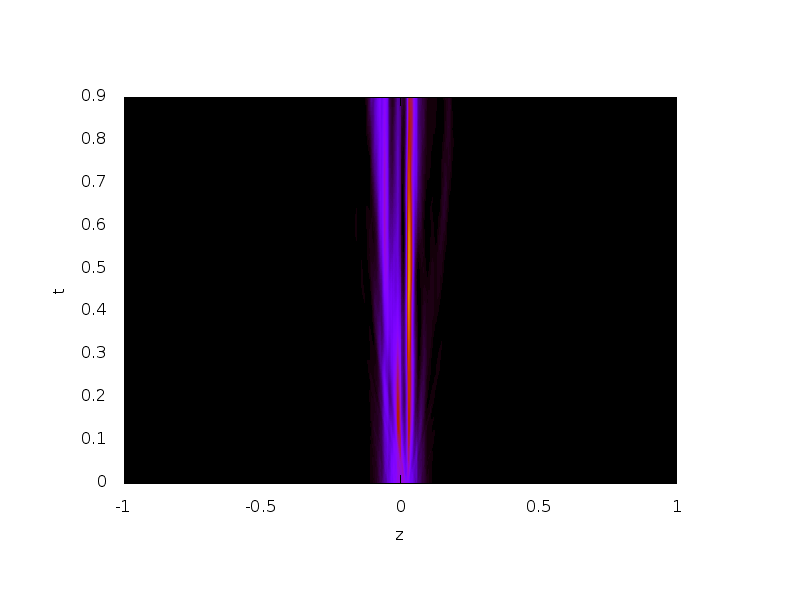}\\
$m_0=20$ \hspace{5cm} $m_0=25$\\ \vspace{0.15cm}
\includegraphics[scale=0.22]{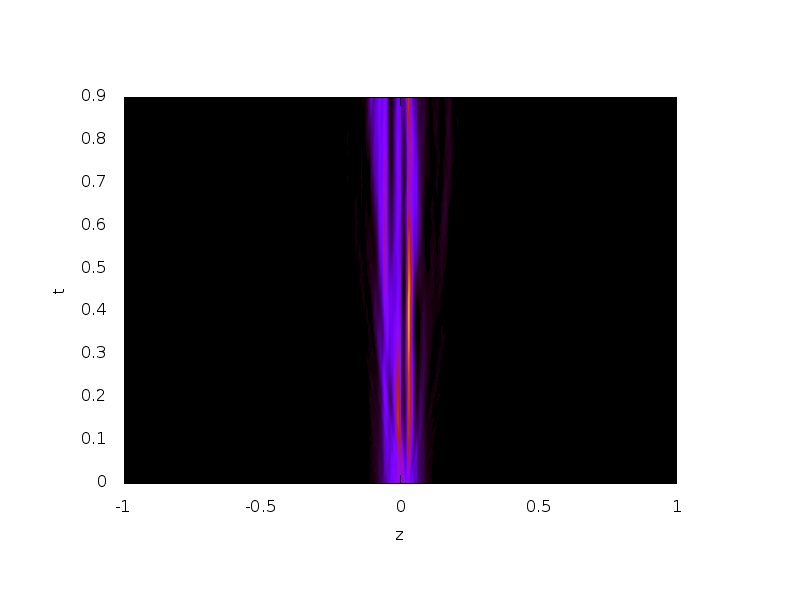}
\includegraphics[scale=0.22]{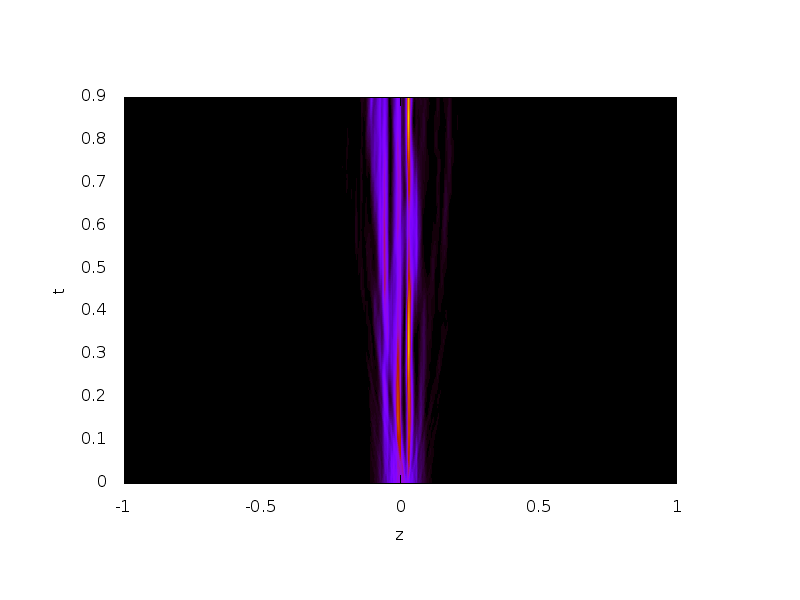}
\caption{$|\psi(z,t)|^2$ as a function of space and time for $\sigma=0.04$, $\overline{m}=500$ and $m_0=0$, $5$, $10$, $15$, $20$ and $25$ from top to bottom and from left to right. Corrugations due to the mass randomness are clearly visible, but they do not imply major qualitative changes of the wavepacket.
}
\label{fig:mass500_psi}
\end{figure}
\begin{figure}[htbp]
\centering
$m_0=0$ \hspace{5cm} $m_0=5$\\ \vspace{0.15cm}
\includegraphics[scale=0.22]{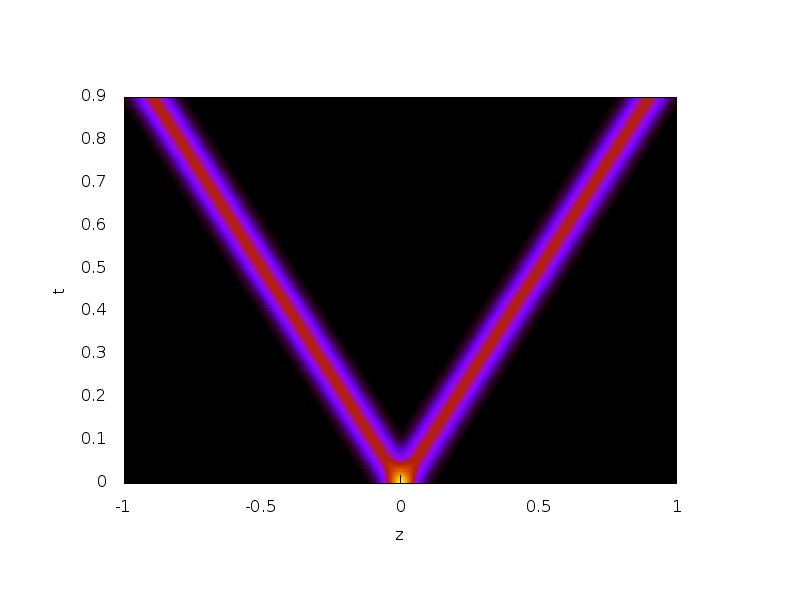}
\includegraphics[scale=0.22]{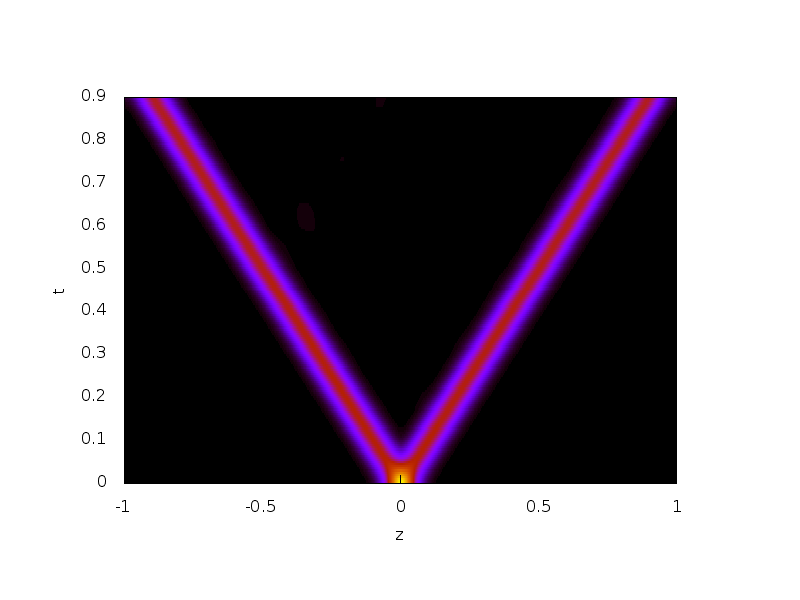}\\
$m_0=10$ \hspace{5cm} $m_0=15$\\ \vspace{0.15cm}
\includegraphics[scale=0.22]{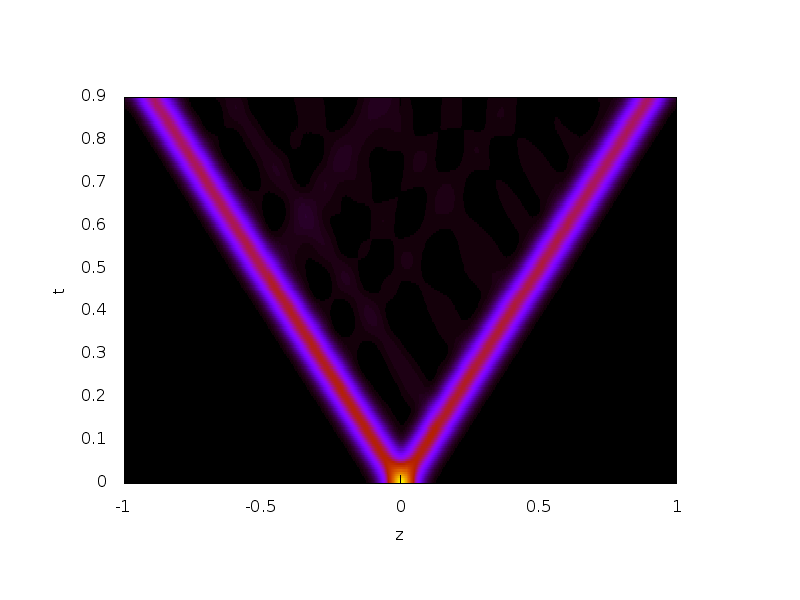}
\includegraphics[scale=0.22]{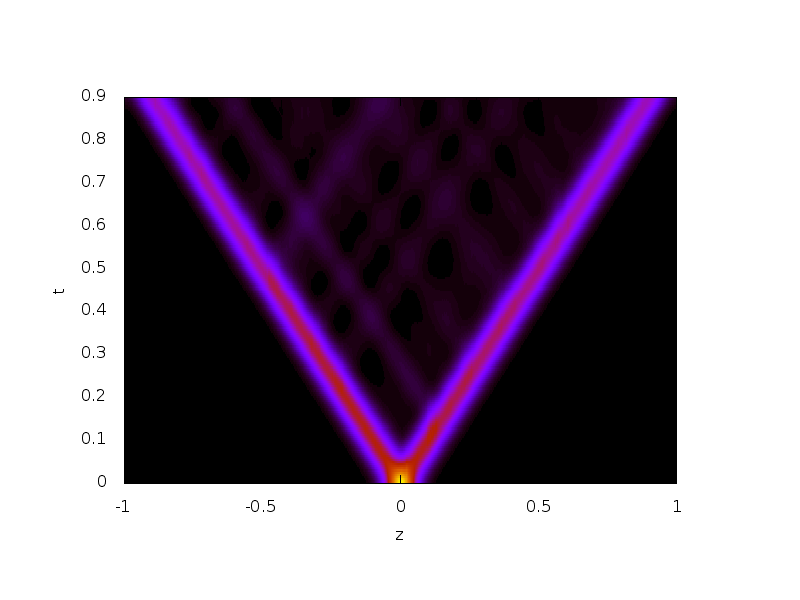}\\
$m_0=20$ \hspace{5cm} $m_0=25$\\ \vspace{0.15cm}
\includegraphics[scale=0.22]{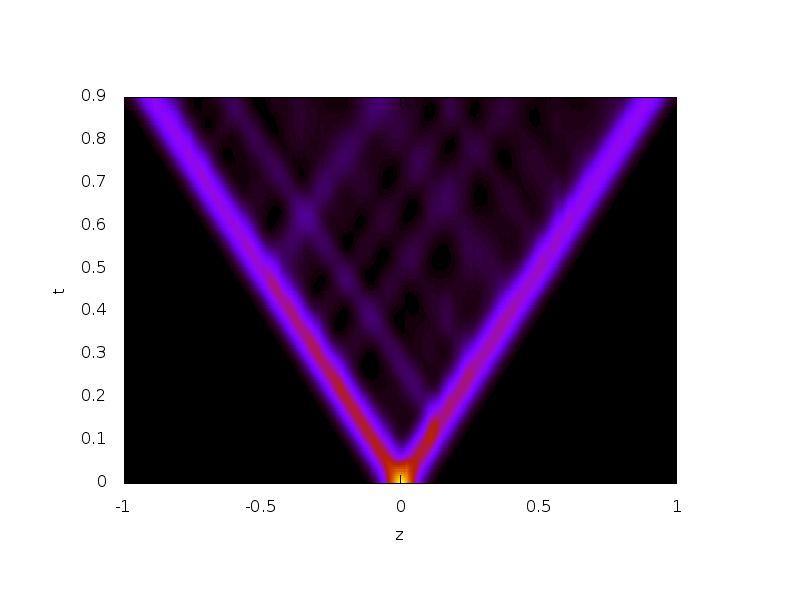}
\includegraphics[scale=0.22]{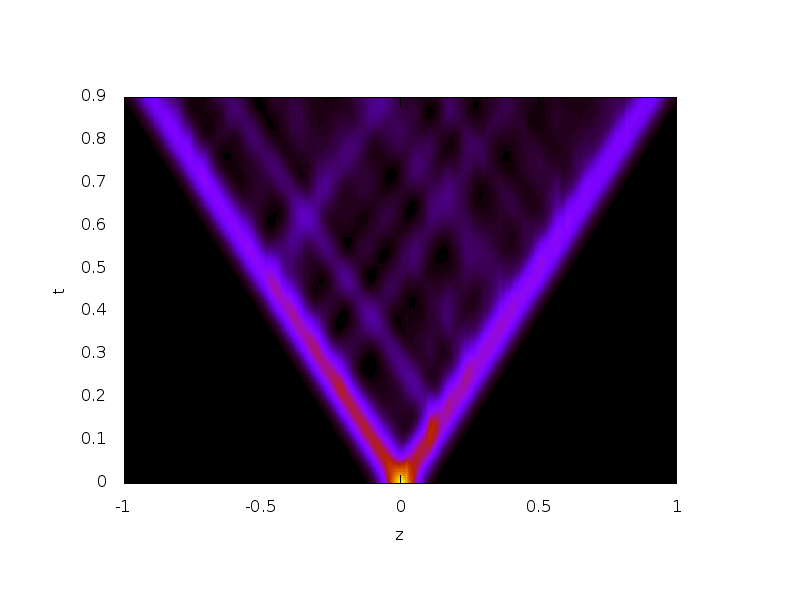}
\caption{$|\psi(z,t)|^2$ as a function of space and time for $\sigma=0.04$, $\overline{m}=0$ and $m_0=0$, $5$, $10$, $15$, $20$ and $25$ from top to bottom and from left to right.
These spacetime representation clearly shows the interference of the different "branches" of the wavefunction.
}
\label{fig:mass0_psi}
\end{figure}

\section{Summary and outlook}

Summarising, we have used the quantum lattice Boltzmann scheme to simulate the one-dimensional
Dirac equation with a random mass. The simulations confirm that a random mass leads to
increasing localization of the wave-function and, for the case of zero-average mass, also to ultra-slow Sinai diffusion.
The latter regime is related to the appearance of a remarkable "networked" structure of the wave
function, due to the onset of a hierarchy of side-branches resulting from fluctuations
causing sign changes of the mass around a zero mean.
These results indicate that the quantum lattice Boltzmann method provides a promising
tool for the numerical simulation of quantum-relativistic transport phenomena in topological materials.

\section{Acknowledgments}
Professor A. Kamenev is kindly acknowledged for critical reading and very useful remarks.

{}


\begin{thebibliography}{99}

\bibitem{BEEN} C. W. J. Beenakker,
Rev. Mod. Phys. 87, 1037 (2015).

\bibitem {MORI}  T. Morimoto, A. Furusaki and C. Mudry,
Phys. Rev. B, 91, 23511 (2015)

\bibitem{PWA} P.W. Anderson,
Phys. Rev. 109, 1492 (1958)

\bibitem{SACH} S. Sachdev, Quantum phase transitions,
Handbook of Magnetism and Advanced Magnetic Materials,
John Wiley and Sons, Ltd  (2007)

\bibitem{KAMEN} D. Bagrets, A. Altland, A. Kamenev,
Sinai Diffusion at Quasi-1D Topological Phase Transitions, Phys. Rev. Lett. 117, 196801, (2016).

\bibitem{QLB1} S. Succi and R. Benzi,
Lattice Boltzmann equation for quantum mechanics, Physica D 69, 327--332 (1993)

\bibitem{SINAI} Y. G. Sinai,
Theory Probab. Appl. 27, 256 (1982).

\bibitem{BOU}
J. Bouchaud, A. Comtet, A. Georges, and P. L. Doussal,
Annals of Physics 201, 285 (1990).

\bibitem{COMTET} A. Comtet and D. S. Dean,
Journal of Physics A: Mathematical and General 31, 8595 (1998).

\bibitem{BAL}  L. Balents and M. P. A. Fisher,
Phys. Rev. B 56, 12970 (1997).

\bibitem{SHANKAR}  R. Shankar and G. Murthy,
Phys. Rev. B 36, 536 (1987).

\bibitem{YOSPRA} A. Yosprakob, S. Suwanna,
arXiv:1601.03827v2[quantum-ph] 6 May 2016, and references therein

\bibitem{LL} L. Landau and E. Lifshitz,
Relativistic Quantum Field Theory, Pergamon, Oxford (1960).

\bibitem{DELLAR1} P. J. Dellar, D. Lapitski, S. Palpacelli and S. Succi, Isotropy of three-dimensional quantum lattice Boltzmann schemes, Phys. Rev. E 83, 046706, (2011).

\bibitem{DELLAR2} D. Lapitski and P. J. Dellar, Convergence of a three-dimensional quantum lattice Boltzmann scheme towards solutions of the Dirac equation, Phil. Trans. R. Soc. Lond. A 369, 2155-2163, (2011).

\bibitem{QLB2} S. Palpacelli, S. Succi,
Quantum lattice Boltzmann simulation of expanding Bose-Einstein condensates in random potentials, Phys. Rev. E 77, 066708 (2008).

\bibitem{QLB3} F. Fillion-Gourdeau, H. J. Herrmann, M. Mendoza, S. Palpacelli, and S. Succi,
Formal Analogy between the Dirac Equation in Its Majorana Form and the Discrete-Velocity Version of the Boltzmann Kinetic Equation,
Phys. Rev. Lett. 111, 160602 (2013).
\end{thebibliography}
\end{document}